\documentclass[preprint2]{aastex61}
\received{Oct 6, 2017}
\revised{Dec 4, 2017}
\accepted{Dec 5, 2017}
\submitjournal{AJ}

\shorttitle{Qatar-6b exoplanet}
\shortauthors{Alsubai et al.}

\usepackage{amsmath,texlogos}
\usepackage{natbib}
\usepackage{subfigure}
\usepackage{graphicx}
\usepackage{txfonts}
\usepackage[T1]{fontenc}
\usepackage{aecompl}
\usepackage{times}

\newcommand{\kms}{\ensuremath{\rm km\,s^{-1}}}
\newcommand{\ms}{\ensuremath{\rm m\,s^{-1}}}
\newcommand{\logg}{\ensuremath{\log{g}}}
\newcommand{\vsini}{\ensuremath{v\,\sin{i}}}

\newcommand{\feh}{\ensuremath{\left[{\rm Fe}/{\rm H}\right]}}
\newcommand{\mh}{\ensuremath{\left[m/{\rm H}\right]}}
\newcommand{\teff}{\ensuremath{T_{\rm eff}}}

\newcommand{\msun}{\ensuremath{\,M_\odot}}
\newcommand{\rsun}{\ensuremath{\,R_\odot}}
\newcommand{\lsun}{\ensuremath{\,L_\odot}}

\newcommand{\mstar}{\ensuremath{\,M_\star}}
\newcommand{\rstar}{\ensuremath{\,R_\star}}
\newcommand{\mpl}{\ensuremath{\,M_{\rm P}}}
\newcommand{\rpl}{\ensuremath{\,R_{\rm P}}}
\newcommand{\mj}{\ensuremath{{\rm \,M_{J}}}}
\newcommand{\rj}{\ensuremath{{\rm \,R_{J}}}}

\begin{document}

\title{Qatar Exoplanet Survey: Qatar-6b -- a grazing transiting hot Jupiter}

\correspondingauthor{Khalid Alsubai}
\email{kalsubai@qf.org.qa}

\author{Khalid Alsubai}
\affil{Qatar Environment and Energy Research Institute (QEERI), HBKU, Qatar Foundation, PO Box 5825, Doha, Qatar}

\author{Zlatan I. Tsvetanov}
\affiliation{Qatar Environment and Energy Research Institute (QEERI), HBKU, Qatar Foundation, PO Box 5825, Doha, Qatar}

\author{David W. Latham}
\affiliation{Harvard-Smithsonian Center for Astrophysics, 60 Garden Street,  Cambridge, MA 02138, USA}

\author{Allyson Bieryla}
\affiliation{Harvard-Smithsonian Center for Astrophysics, 60 Garden Street,  Cambridge, MA 02138, USA}

\author{Gilbert A. Esquerdo}
\affiliation{Harvard-Smithsonian Center for Astrophysics, 60 Garden Street,  Cambridge, MA 02138, USA}

\author{Dimitris Mislis}
\affiliation{Qatar Environment and Energy Research Institute (QEERI), HBKU, Qatar Foundation, PO Box 5825, Doha, Qatar}

\author{Stylianos Pyrzas}
\affiliation{Qatar Environment and Energy Research Institute (QEERI), HBKU, Qatar Foundation, PO Box 5825, Doha, Qatar}



\author{Emma Foxell}
\affiliation{Department of Physics, University of Warwick, Gibbet Hill Road, Coventry CV4 7AL, UK}
\affiliation{Centre for Exoplanets and Habitability, University of Warwick, Gibbet Hill Road, Coventry CV4 7AL, UK}

\author{James McCormac}
\affiliation{Department of Physics, University of Warwick, Gibbet Hill Road, Coventry CV4 7AL, UK}
\affiliation{Centre for Exoplanets and Habitability, University of Warwick, Gibbet Hill Road, Coventry CV4 7AL, UK}

\author{Christoph Baranec}
\affiliation{Institute for Astronomy, University of Hawai`i at M\={a}noa, Hilo, HI 96720-2700, USA}


\author{Nicolas P. E. Vilchez}
\affiliation{Qatar Environment and Energy Research Institute (QEERI), HBKU, Qatar Foundation, PO Box 5825, Doha, Qatar}

\author{Richard West}
\affiliation{Department of Physics, University of Warwick, Gibbet Hill Road, Coventry CV4 7AL, UK}
\affiliation{Centre for Exoplanets and Habitability, University of Warwick, Gibbet Hill Road, Coventry CV4 7AL, UK}

\author{Ali Esamdin}
\affiliation{Xinjiang Astronomical Observatory, Chinese Academy of Sciences, 150 Science 1-Street, Urumqi, Xinjiang 830011, China }

\author{Zhenwei Dang}
\affiliation{Xinjiang Astronomical Observatory, Chinese Academy of Sciences, 150 Science 1-Street, Urumqi, Xinjiang 830011, China }

\author{Hani M. Dalee}
\affiliation{Qatar Environment and Energy Research Institute (QEERI), HBKU, Qatar Foundation, PO Box 5825, Doha, Qatar}

\author{Amani A. Al-Rajihi}
\affiliation{Qatar Secondary Independent High School, Doha, Qatar}

\author{Abeer Kh. Al-Harbi}
\affiliation{Al-Kawthar Secondary Independent High School, Doha, Qatar}




\begin{abstract}
We report the discovery of Qatar-6b, a new transiting planet identified by the Qatar Exoplanet 
Survey (QES). The planet orbits a relatively bright (V=11.44), early-K main-sequence star at 
an orbital period of $P\sim3.506$ days. An SED fit to available multi-band photometry, ranging 
from the near-UV to the mid-IR, yields a distance of $d = 101\,\pm\,6$ pc to the system. From 
a global fit to follow-up photometric and spectroscopic observations, we calculate the mass 
and radius of the planet to be \mpl\,=\,0.67$\pm$0.07\mj\ and \rpl\,=\,1.06$\pm0.07$\rj\ 
respectively. We use multi-color photometric light curves to show that the transit is grazing, 
making Qatar-6b one of the few exoplanets known in a grazing transit configuration. It adds 
to the short list of targets that offer the best opportunity to look for additional bodies in the 
host planetary system through variations in the transit impact factor and duration. 

\end{abstract}





\section{Introduction}

Dedicated ground-based photometric surveys continue to provide a steady supply of 
short period, transiting exoplanets orbiting relatively bright stars. Subsequent studies 
of hot Jupiters and other close-orbiting large planets -- a type of planets not present in 
our solar system -- contribute key information towards understanding the overall 
structure and composition of planetary systems.

Over the past two decades, transiting exoplanets have been the subject of a number 
of different studies, yielding a wide range of results, including identifying the presence 
of additional bodies using transit-time variations (e.g., \citealt{agol}, \citealt{holmur});
detection of thermal emission (e.g., \citealt{TrES-1}, \citealt{deming05}, \citealt{deming06}); 
spin-orbit alignment or lack thereof (e.g., \citealt{albrecht}, see also \citealt{wifa} for a 
review) and studies of exoplanets' atmospheres (e.g., \citealt{HD209458}, \citealt{tinetti}, 
\citealt{sing}) to name but a few.

Although space-based surveys have dramatically increased the number of (fully)
transiting exoplanets, \emph{grazing} (or nearly- so) transiting exoplanets still
constitute only a tiny fraction of the known exoplanet population. In fact, to the 
best of our knowledge, their number has only recently reached double-digits; 
systems identified by ground-based surveys are WASP-34b \citep{wasp34}; 
WASP-45b \citep{wasp45}; WASP-67b \citep{wasp67}; WASP-140b \citep{wasp140} 
and HAT-P-27b/WASP-40b (\citealt{hat27}/\citealt{wasp40}), coupled with the 
space-based detections of CoRoT-25b \citep{corot25}; CoRoT-33b \citep{corot33};
Kepler-434b \citep{kepler434}; Kepler-447b \citep{kepler447} and K2-31b \citep{k2-31} 
to complete the, rather small, family of ten.

Planets in grazing transit configurations offer an intriguing, yet hitherto untested, 
avenue of study \citep{kepler447}. In short, any external gravitational influence on 
the system, e.g., the presence of additional bodies such as an outer planet or even 
an exo-moon, would perturb the grazing planet's orbit and could potentially induce 
periodic variations of the transit impact parameter (TIP, \citealt{kipping09}), leading 
to transit duration variations (TDV, \citealt{kipping10}). With sufficient cadence and 
photometric accuracy, these variations would be detectable in the transit light curve, 
and could be used in turn to study the perturbing body.

In this paper we present the discovery of Qatar-6b, a newly found hot Jupiter 
on a grazing transit. The paper is organized as follows: in Section 
\ref{sec:Observations} we present the survey photometry and describe the 
follow-up spectroscopy and photometry used to confirm the planetary nature 
of the transits. In Section \ref{sec:Analysis} we present analysis of the data and 
the global system solutions using simultaneous fits to the available RV and 
follow-up photometric light curves, and in Section \ref{sec:conclusions} we 
summarize the results. 

\section{Observations} \label{sec:Observations} 

\subsection{Discovery photometry} \label{subsec:DiscPhot}

The survey data were collected with the Qatar Exoplanet Survey (QES) hosted by the 
New Mexico Skies Observatory\footnote{http://www.nmskies.com} located at Mayhill, 
NM, USA. A full description of QES can be found in our previous publications, e.g., 
\cite{alsubai2013}, \cite{alsubai2017}. For completeness, here we give the main 
survey characteristics. QES uses two overlapping wide field 135\,mm (f/2.0) and 
200\,mm (f/2.0) telephoto lenses, along with four 400\,mm (f/2.8) telephoto lenses, 
mosaiced to image an $11{\degr}\,\times\,11{\degr}$ field on the sky simultaneously 
at three different pixel scales --- 12, 9 and 4 arcsec, respectively, for the three 
different focal length lenses. All lenses are equipped with FLI ProLine PL6801 
cameras, with KAF-1680E 4k$\times$4k detectors. Exposure times are 60\,s, 
for each of the four CCDs attached to the 400\,mm lenses; 45\,s, for the CCD 
equipped with the 200\,mm lens; and 30\,s, for the CCD equipped with the 
135\,mm lens.The combination of large aperture lenses and high survey angular 
resolution, allow QES to reach 1\% photometric precision up to 13.5-14.0 mag 
stars. 

The data were reduced with the QES pipeline, which performs bias-correction, 
dark-current subtraction and flat-fielding in the standard fashion, while photometric 
measurements are extracted using the image subtraction algorithm by \cite{dbdia}; 
a more detailed description of the pipeline is given in \cite{alsubai2013}.

\begin{figure}
\centering
\includegraphics[width=8.5cm]{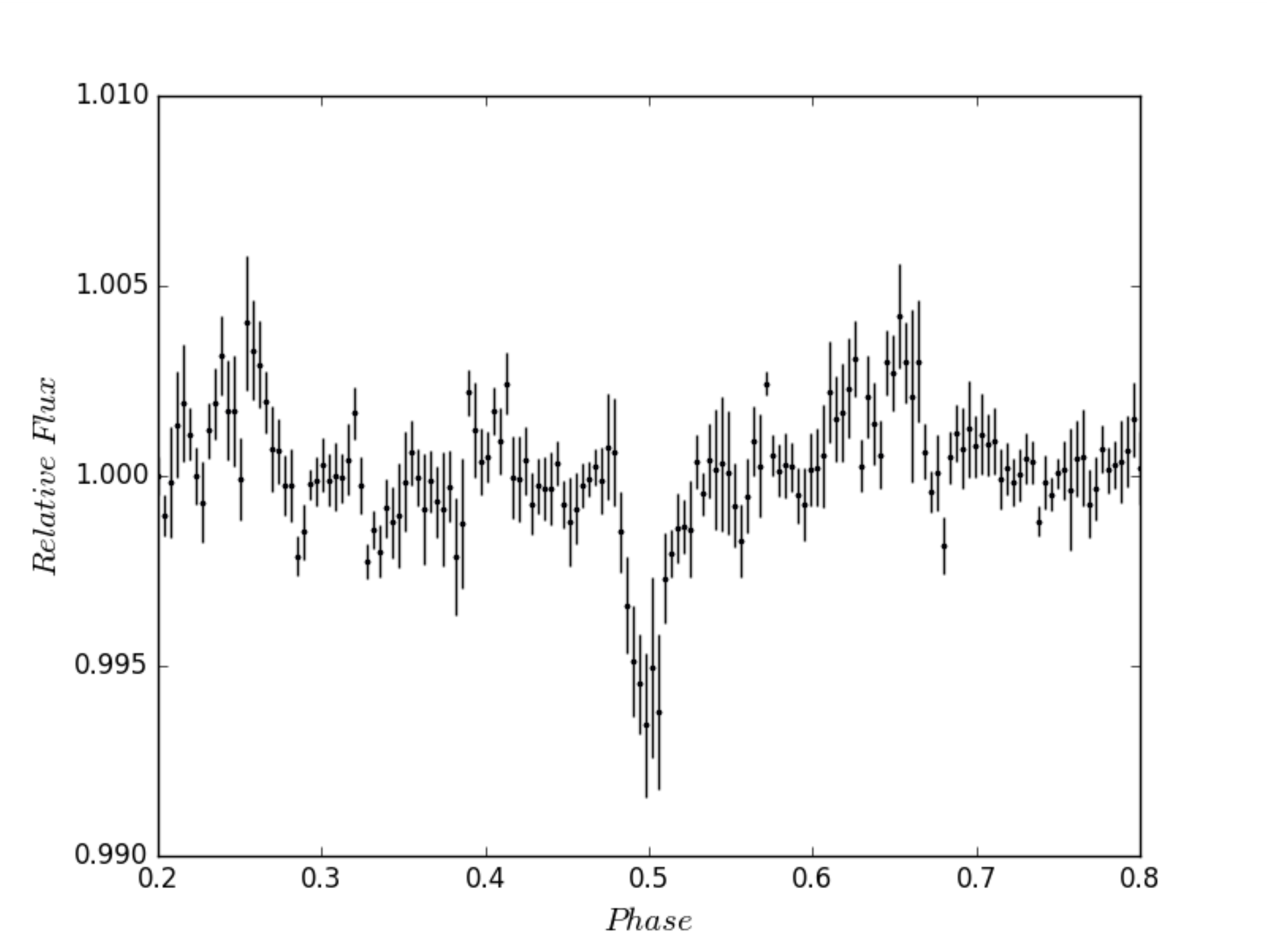} 
\caption{The discovery light curve for Qatar-6b phase folded with the BLS estimated 
period, as it appears in the QES archive.}
\label{discoveryLC}
\end{figure}

The output light curves are ingested into the QES archive and subjected to a 
combination of the Trend Filtering Algorithm (TFA, \citealt{kovacs1}) and the Doha 
algorithm \citep{mislis}, to model and remove systematic patterns of correlated noise. 
Transit-like events are identified using the Box Least Squares algorithm (BLS) of 
\cite{kovacs2}, during a candidates search on the archive light curves following the 
procedure described in \cite{collier}. We note that the initial candidate selection is 
an automatic procedure, but the final vetting is done by eye. The BLS algorithm 
provided a tentative ephemeris which was used to phase-fold the discovery light 
curve shown in Figure\,\ref{discoveryLC}. The discovery light curve of Qatar-6b 
contains 2324 data points obtained from March to July 2012. 

The host of Qatar-6b is a $V$\,=\,11.44 mag ($B$\,=\,12.34 mag) high proper motion 
star (TYC 1484-434-1, 2MASS J14485047+2209093, henceforth designated Qatar-6) 
of spectral type close to K2V. A detailed discussion of stellar parameters based on the 
analysis of our follow-up spectra and on the available photometry is presented in 
Section \ref{subsec:SpecPars}. Here we note only that the host star spectral type is 
initially estimated from a multi-color ($V$, $J$, $H$ and $K$ bands) fit to the 
magnitudes, using a standard Random-Forest classification algorithm, trained with 
$\sim$2000 standards with spectral types ranging from early A to late M.

\subsection{Follow-up photometry} \label{subsec:FollowPhot}

Follow-up photometric observations for Qatar-6b were collected with the 1.2\,m telescope 
at the Fred L.\,Whipple Observatory (FLWO, Mount Hopkins, Arizona) using KeplerCam, a 
single 4K $\times$ 4K CCD that covers an area of 23$^{\prime} \times 23^{\prime}$ on the 
sky. The target was observed on four occasions -- (1) on the night of January 30, 2017 
through Sloan $i$ filter; (2) on May 5, 2017 (Sloan $i$); (3) on May 19, 2017 (Sloan $r$); 
and (4) on June 9, 2017 (Sloan $z$). 

Two additional transits were obtained using the Near Infra-red Transiting ExoplanetS 
telescope \citep[NITES,][]{McCormac14} located at the Observatorio del Roque de los 
Muchachos (ORM, La Palma, Canary Islands), on March 9, 2017 and on March 23, 2017. 
On both nights, a total of 354 30\,s images were obtained with a Johnson-Bessel $V$-band 
filter. The data were reduced in {\scshape Python} using {\scshape ccdproc} \citep{Craig15}. 
A master bias, dark and flat was created using the standard process on each night. A 
minimum of 21 of each frame was used in each master calibration frame. Non-variable 
nearby comparison stars were selected by hand and aperture photometry extracted using 
{\scshape sep} \citep{Barbary16, Bertin96}. 

\begin{figure*}
\centering
\includegraphics[width=8.8cm]{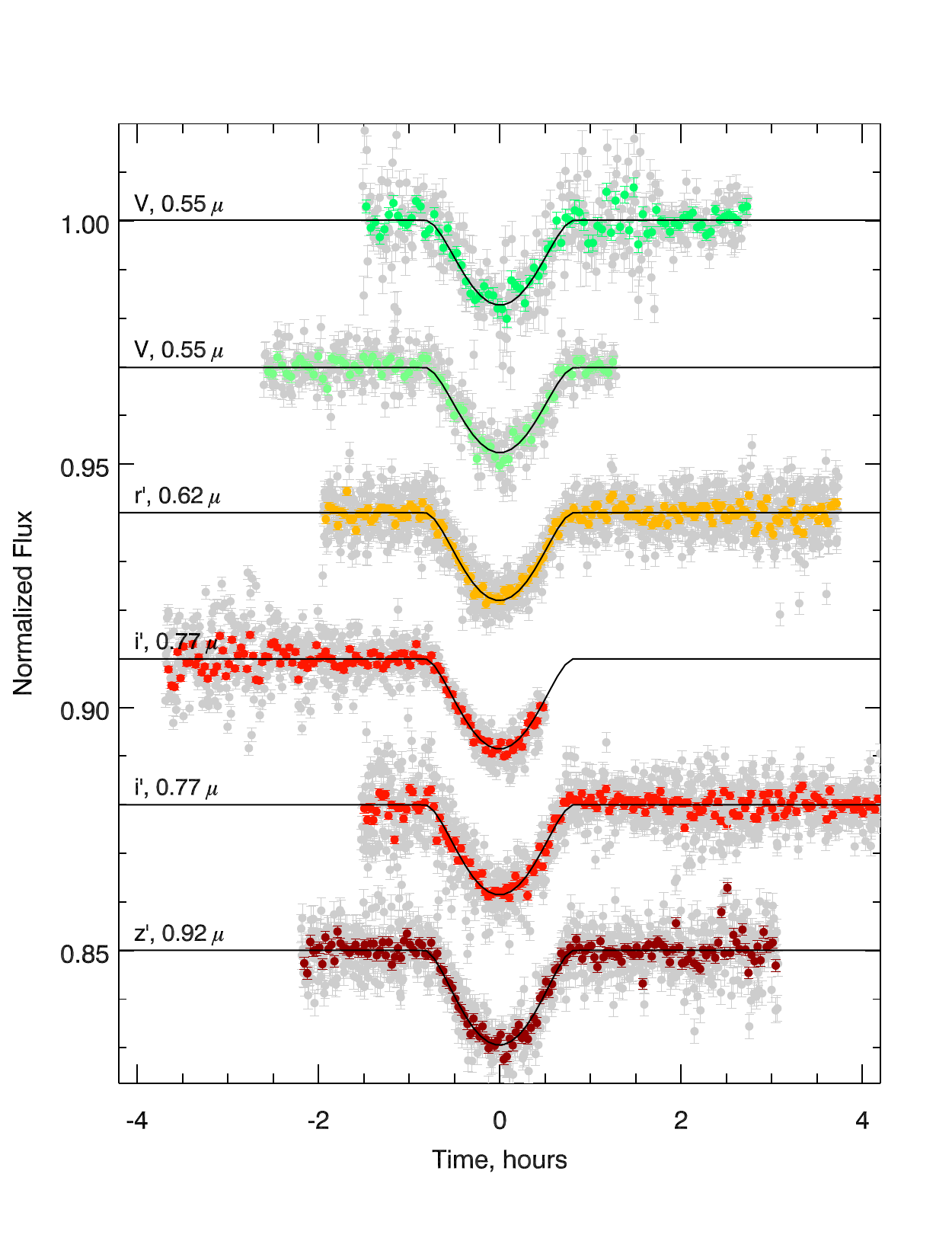}
\includegraphics[width=8.8cm]{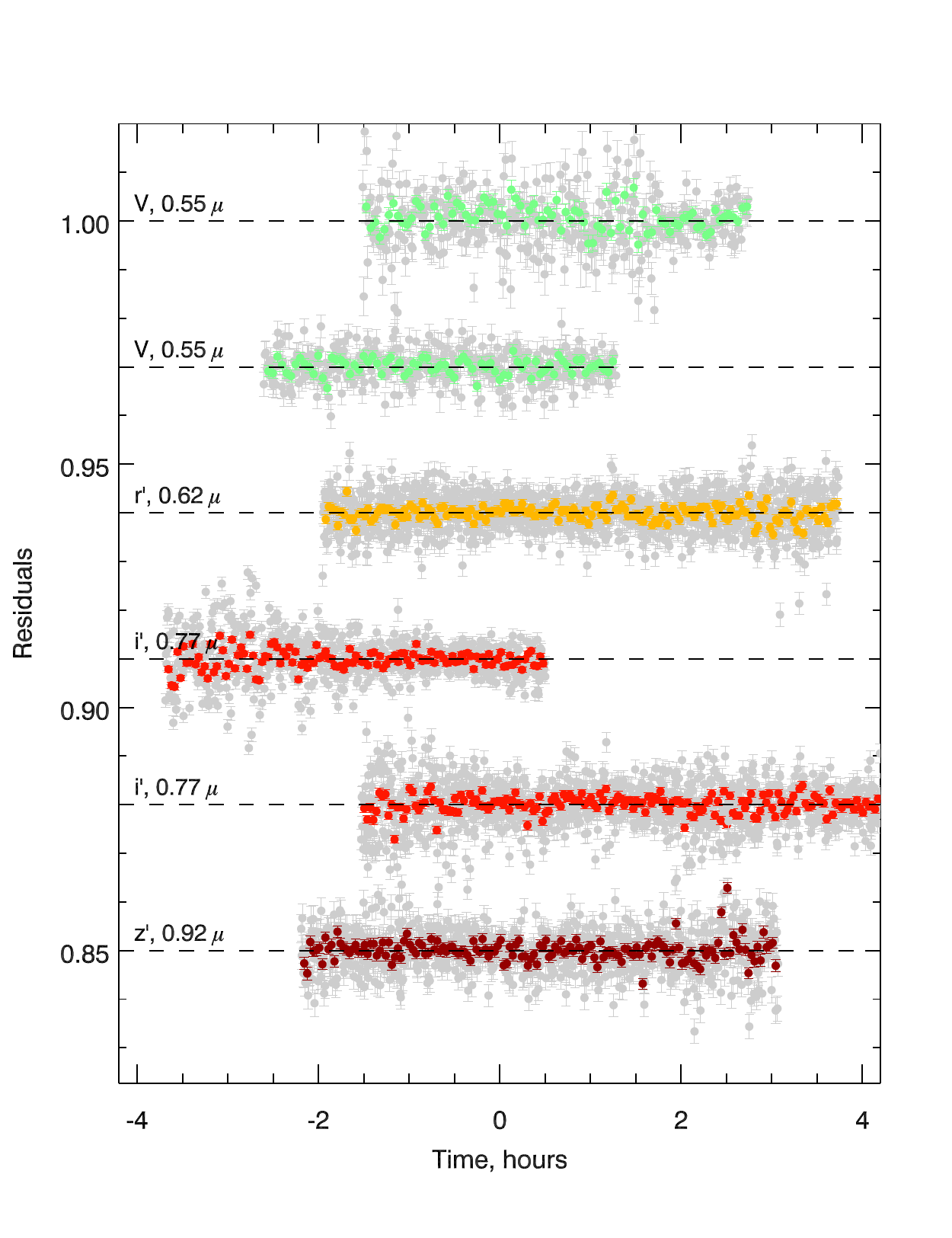}
\caption{The six follow-up light curves of Qatar-6b. The {\it left} panel shows the light 
curves ordered from top to bottom in increasing filter wavelength (the light curves
have been shifted vertically for clarity). The original observational data points are 
plotted in light gray and binned data are shown in darker colors (see Section 
\ref{subsec:OrbitalPeriod}). The solid, black lines represent the best model fit for the 
corresponding filter. The residuals from the fits are shown in the {\it right} panel. The 
dates of observations for each light curve are given in the text.}
\label{fig:Q6bTR}
\end{figure*}

Figure\,\ref{fig:Q6bTR} shows the follow-up light curves together with the model fits described 
in Section \ref{subsec:EXOFAST}. 


\subsection{Follow-up spectroscopy} \label{subsec:FollowSpec}

Similar to our campaigns for all QES candidates, follow-up spectroscopic observations 
were obtained with the Tillinghast Reflector Echelle Spectrograph (TRES) on the 
1.5\,m Tillinghast Reflector at the FLWO. We used TRES in a configuration with the 
medium fiber, which yields a resolving power of $R \sim$ 44,000, corresponding to 
a velocity resolution element of 6.8 \kms\ FWHM. The spectra were extracted using 
version 2.55 of the code described in \cite{buchhave2010}. The wavelength calibration 
for each spectrum was established using exposures of a thorium-argon hollow-cathode 
lamp illuminating the science fiber, obtained immediately before and after each 
observation of the star. 

\begin{figure}
\centering
\includegraphics[width=8.5cm]{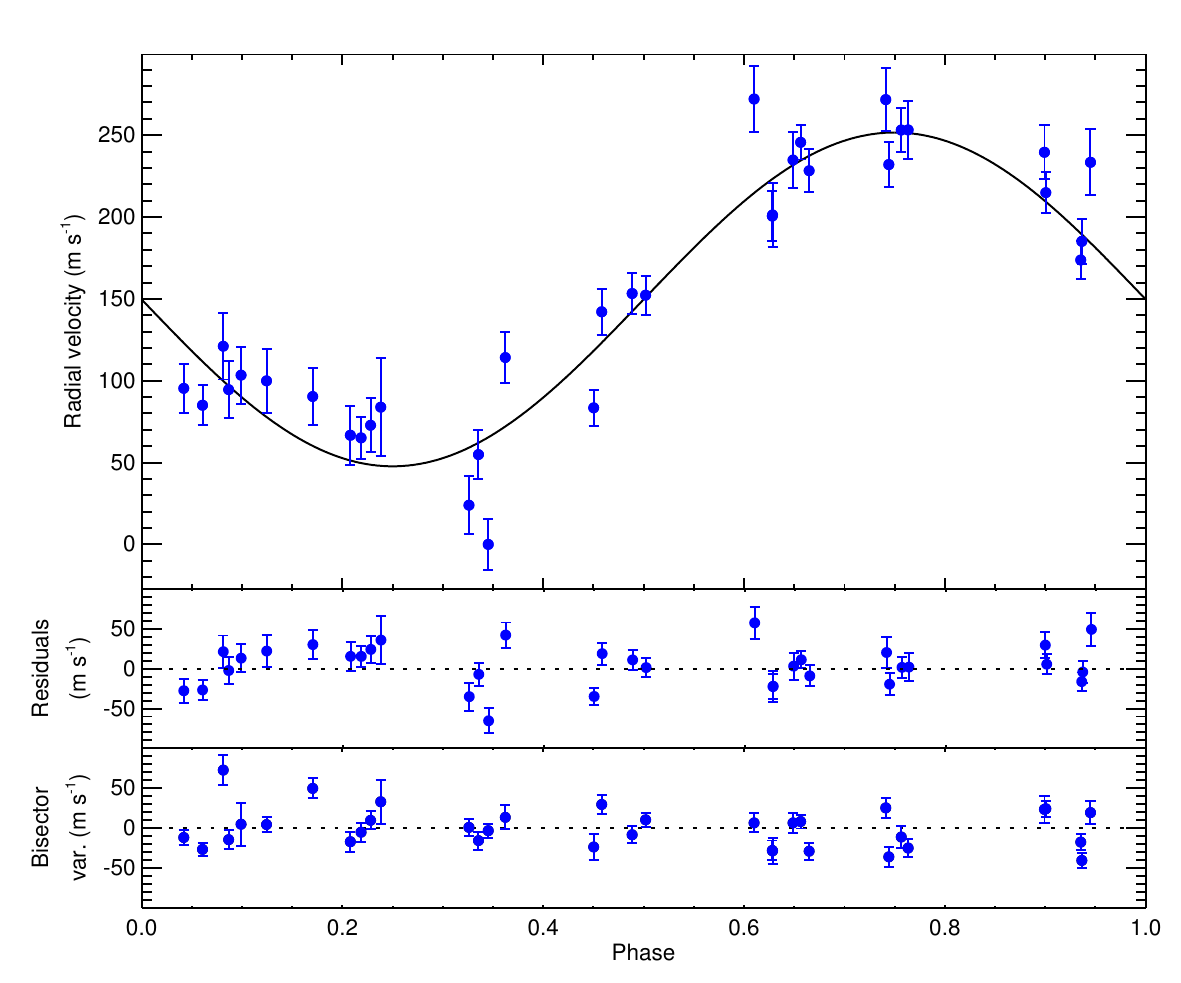}
\caption{Orbital solution for Qatar-6b, showing the velocity curve and observed 
velocities and the bisector values.}
\label{figureQ6bRV}
\end{figure}

For Qatar-6 a total of 34 spectra were obtained between April 13 -- May 31, 2016 
with exposure times ranging from 450 sec to 1600 sec and an average signal-to-noise 
ratio per resolution element (SNRe) of $\sim$36 at the peak of the continuum in the 
echelle order centered on the Mg b triplet near 519 nm. Relative radial velocities (RV) 
were derived by cross-correlating each observed spectrum against the strongest 
exposure of the same star, order by order for a set of echelle orders selected to have 
good SNRe and minimal contamination by telluric lines introduced by the Earth's 
atmosphere. These RVs are reported in Table \ref{table:RV} (with the time values in 
Barycentric Julian Date in Barycentric Dynamical time, BJD$_{TDB}$) and plotted in 
Figure\,\ref{figureQ6bRV}. The observation that was used for the template spectrum 
has, by definition, a RV of 0.0 \kms. The error on the template RV is defined as the 
smallest error of all the other errors. We also derived values for the line profile bisector 
spans (BS, lower panel in Figure \ref{figureQ6bRV}), to check for astrophysical 
phenomena other than orbital motion that might produce a periodic signal in the 
RVs with the same period as the photometric ephemeris for the transits. The 
procedures used to determine RVs and BSs are outlined in \citet{buchhave2010}.

\begin{table}
\centering
\caption{Relative RVs and BS variations for Qatar-6.}
\label{table:RV}
\begin{tabular}{ccc}
\hline
BJD$_{TDB}$       &RV (\ms)          &BS (\ms)       \\
\hline
$ 2457491.849996 $ & $ 228 \pm 13 $ & $ -29 \pm 11 $ \\
$ 2457496.967581 $ & $ 100 \pm 20 $ & $   4 \pm 10 $ \\
$ 2457498.832531 $ & $ 246 \pm 10 $ & $   8 \pm  9 $ \\
$ 2457499.810664 $ & $ 174 \pm 11 $ & $ -18 \pm 10 $ \\
$ 2457500.837013 $ & $  73 \pm 17 $ & $  10 \pm 11 $ \\
$ 2457501.797354 $ & $ 152 \pm 12 $ & $ 10 \pm  9 $ \\
$ 2457503.847470 $ & $  95 \pm 17 $ & $ -15 \pm 12 $ \\
$ 2457504.813571 $ & $ 114 \pm 16 $ & $  14 \pm 15 $ \\
$ 2457505.818142 $ & $ 234 \pm 17 $ & $   7 \pm 13 $ \\
$ 2457506.856937 $ & $ 233 \pm 20 $ & $ 19 \pm 15 $ \\
$ 2457507.778682 $ & $  67 \pm 18 $ & $ -17 \pm 13 $ \\
$ 2457508.762515 $ & $ 153 \pm 12 $ & $  -8 \pm 11 $ \\
$ 2457509.702250 $ & $ 253 \pm 13 $ & $ -11\pm 14 $ \\
$ 2457510.768550 $ & $  85 \pm 12 $ & $ -27 \pm  9  $ \\
$ 2457511.766238 $ & $   0 \pm 16 $ & $   -3 \pm 9 $ \\
$ 2457512.758133 $ & $ 201 \pm 15 $ & $ -28 \pm 13 $ \\
$ 2457513.838902 $ & $ 185 \pm 14 $ & $ -41 \pm  9 $ \\
$ 2457514.828265 $ & $  65 \pm 13 $ & $  -5 \pm 12 $ \\
$ 2457523.750772 $ & $ 253 \pm 18 $ & $ -25 \pm 12 $ \\
$ 2457524.927611 $ & $ 103 \pm 17 $ & $   5 \pm 27 $ \\
$ 2457526.718970 $ & $ 272 \pm 20 $ & $   7 \pm 12 $ \\
$ 2457527.732882 $ & $ 240 \pm 17 $ & $ 24 \pm 17 $ \\
$ 2457528.683977 $ & $  90 \pm 18 $ & $ 50 \pm 13 $ \\
$ 2457529.693135 $ & $ 142 \pm 14 $ & $ 30 \pm 12 $ \\
$ 2457530.685060 $ & $ 272 \pm 19 $ & $ 25 \pm 12 $ \\
$ 2457531.877658 $ & $ 121 \pm 20 $ & $ 73 \pm 18 $ \\
$ 2457532.735667 $ & $  24 \pm 18 $ & $   1 \pm 11 $ \\
$ 2457533.795727 $ & $ 201 \pm 20 $ & $ -29 \pm 16 $ \\
$ 2457534.750027 $ & $ 215 \pm 12 $ & $  24 \pm 10 $ \\
$ 2457535.933783 $ & $  84 \pm 30 $ & $  33 \pm 28 $ \\
$ 2457536.677373 $ & $  83 \pm 11 $ & $ -24 \pm 16 $ \\
$ 2457537.707875 $ & $ 232 \pm 14 $ & $ -36 \pm 13 $ \\
$ 2457538.752114 $ & $  95 \pm 15 $ & $ -12 \pm  9 $ \\
$ 2457539.781151 $ & $  55 \pm 15 $ & $ -16 \pm 11 $ \\
\hline
\end{tabular}
\end{table}

To get the absolute center-of-mass velocity for the system ($\gamma$), we have 
to provide an absolute velocity for the observation that was used for the template 
when deriving the relative velocities. To derive an absolute velocity for that 
observation, we correlate the Mg b order against the template from the CfA library 
of synthetic templates that gives the highest peak correlation value. Then we add 
the relative $\gamma$-velocity from the orbital solution, and also correct by 
$-0.61$ \kms, mostly because the CfA library does not include the gravitational 
redshift. This offset has been determined empirically by many observations of 
IAU Radial Velocity Standard Stars. We quote an uncertainty in the resulting 
absolute velocity of $\pm 0.1$ \kms, which is an estimate of the residual 
systematic errors in the IAU Radial Velocity Standard Star system.

\section{Analysis and Results} \label{sec:Analysis} 


\subsection{Host star spectroscopic parameters} \label{subsec:SpecPars}

To derive the host star characteristics, we analyzed the TRES spectra using the 
Stellar Parameter Classification (SPC) tool developed by \cite{buchhave2012}. The 
SPC cross correlates the observed spectrum with a library of synthetic spectra from 
Kurucz model atmospheres and finds the stellar parameters from a multi-dimensional 
surface fit to the peak correlation values. We used the ATLAS9 grid of models with 
the Opacity Distribution Functions from \cite{ODF}. The SPC determined stellar 
parameters --- effective temperature (\teff), metallicity (\mh\footnote{Note: The SPC 
determines metallicity as \mh, i.e., the total metal content is adjusted while individual 
elements abundances are kept fixed at the solar ratios.}), surface gravity 
(\logg), and projected rotational velocity \vsini\ --- are given in Table 
\ref{tab:StarPars}. These values are calculated by averaging the stellar 
parameters derived for each spectrum individually and weighted by the height 
of the cross-correlation function.  \teff, \logg, and \mh\ are then used as input 
parameters for the Torres relations (\citealt{torres}) to derive estimates for the 
stellar mass and radius, yielding \mstar\,= 0.820\,\msun \,and \rstar\,=\,0.714\,\rsun. 

 

\begin{table*}
\centering
\caption{\label{tab:StarPars} Basic observational parameters 
of Qatar-6b host star and photometry used for the SED fit.}
\begin{tabular}{lllll}
\hline \hline
Parameter & Description & Value  & Source & Ref. \\ 
\hline
\multicolumn{2}{l}{Names} & & & \\
 & \multicolumn{3}{l}{225-118642 (UCAC3), TYC 1484-434-1}  & \\
 & \multicolumn{3}{l}{SDSS J144850.45+220909.2, 2MASS J14485047+2209093}   & \\
 & \multicolumn{3}{l}{GALEX J14485047+2209091, WISE J14485047+2209091}   & \\
\multicolumn{2}{l}{Astrometry} & & & \\
$\alpha_{\mathrm{2000}}$ & RA (J2000) & $14^{\mathrm{h}}48^{\mathrm{m}}50.42^{\mathrm{s}}$ & GAIA &1 \\    
$\delta_{\mathrm{2000}}$  & DEC (J2000) & $+22^{\mathrm{o}}09^{\prime}09.40^{\prime\prime}$ & GAIA & 1 \\
$\mu_{\alpha}$ & Proper motion, R.A., mas\,yr$^{-1}$ & -51.9 $\pm$ 1.1 & GAIA & 1 \\
$\mu_{\delta}$ & Proper motion, DEC, mas\,yr$^{-1}$ & 14.8 $\pm$ 1.1 & GAIA & 1 \\
\multicolumn{2}{l}{Photometry} & & & \\
NUV & GALEX NUV, mag & 18.65 $\pm$ 0.03 & GALEX & 2 \\   
$B$ & Johnson $B$, mag & 12.389$\pm$0.046 & APASS & 7 \\  
$V$ & Johnson $V$, mag & 11.438$\pm$0.080  & APASS & 7 \\ 
$u$ & Sloan $u$, mag & 13.892 $\pm$ 0.02 & SDSS & 3 \\       
$g$ & Sloan $g$, mag & 11.845 $\pm$ 0.02 & SDSS & 4 \\       
$r$ & Sloan $r$, mag & 11.070 $\pm$ 0.03 & this work & \\       
$i$ & Sloan $i$, mag & 10.91 $\pm$ 0.05 & this work & \\          
$z$ & Sloan $z$, mag & 10.76 $\pm$ 0.05 & this work & \\        
$J$  & 2MASS $J$, mag & 9.711 $\pm$ 0.028 & 2MASS & 5 \\   
$H$ & 2MASS $H$, mag & 9.307 & 2MASS & 5 \\                       
$Ks$ & 2MASS $Ks$, mag & 9.225 & 2MASS & 5 \\                    
W1 & WISE1, mag & 9.018 $\pm$ 0.022 & WISE & 6 \\               
W2 & WISE2, mag & 9.078 $\pm$ 0.020 & WISE & 6\\                
W3 & WISE3, mag & 9.059 $\pm$ 0.022 & WISE & 6 \\               
W4 & WISE4,  mag & 9.405 $\pm$ 0.405 & WISE & 6 \\              
\multicolumn{2}{l}{Spectroscopic parameters } & & & \\
       & Spectral type & K2V & this work & \\
$\gamma_{\rm abs}$ &Systemic velocity, \kms\ & $-27.864\pm0.1$ & this work & \\
\vsini & Rotational velocity, \kms & 2.9$\pm$0.5 & this work & \\
$P_{rot}$ &Rotation period, days & $12.75\pm1.75$ & this work &\\
$\tau$ & Age, Gyr & $1.0\pm0.5$ & this work & \\
$A_{V}$ & Extinction, mag & $0.093\pm0.003$ & S\&F2011 & 8 \\
$d$ & Distance, pc & $101\pm6$ & this work & \\
    \hline
\end{tabular}
\tablerefs{(1) GAIA DR1 \href{http://gea.esac.esa.int/archive/}{http://gea.esac.esa.int/archive/}, 
	(2) GALEX \href{http://galex.stsci.edu/}{http://galex.stsci.edu/}, (3) \cite{Pickles2010}, 
	(4) SDSS DR13 \cite{SDSS-DR13}, 
	(5) 2MASS \href{http://irsa.ipac.caltech.edu/Missions/2mass.html}{http://irsa.ipac.caltech.edu/Missions/2mass.html}, 
	(6) WISE \href{http://irsa.ipac.caltech.edu/Missions/wise.html}{http://irsa.ipac.caltech.edu/Missions/wise.html},
	(7) APASS \href{https://www.aavso.org/apass}{https://www.aavso.org/apass},
	(8) \cite{Extinction}}
\end{table*}

The host star age is an important parameter for understanding the evolution of exoplanetary 
systems. We estimate the age of Qatar-6 using both the gyrochronology and isochrone fitting 
methods. For the gyrochronology age, we followed the formalism of \cite{brown} assuming 
the stellar rotation axis is perpendicular to the orbital plane. There are four equations in 
Brown's work describing the rotational period-color-age relation --- in our case, equations 
2, 3, and 4 give consistent estimates of the star's age, while eq.\,1 (based on the $B-V$ 
color, see Section \ref{subsec:SEDfit}) gives about half the value. The uncertainty of the 
age estimate is entirely dominated by the errors in determining the stellar rotation 
($\pm$0.5 \kms) leading to age estimates, $\tau_{\rm gyr}$, in the range 0.75--1.5 Gyr 
(eq.\ 2, 3 and 4) and 0.4-0.6 Gyr (eq.\,1). 

Additionally, using the model isochrones from the Dartmouth database (\citealt{dotter}) 
and input parameters from Table \ref{table:ExoQ6}, we calculate an independent value 
for the age of the host star as $\tau_{\rm iso}  \approx 1.0$ Gyr. Furthermore, the global 
solution for the system parameters presented in Section \ref{subsec:EXOFAST}, which 
uses the YY isochrones (\citealt{YY}), gives a star age of $\tau_{\rm iso} = 1.02 \pm 0.62$ 
Gyr.  In Table \ref{tab:StarPars} we quote the age of the host star as 1 Gyr, which is where 
all our estimates converge and put the uncertainty conservatively at 0.5 Gyr.

\subsection{SED fit and distance determination} \label{subsec:SEDfit}

The host star of Qatar-6b has been observed by a number of imaging surveys covering 
the entire wavelength range 0.23 -- 22\,$\mu$m, i.e. from the GALEX NUV to the WISE 
IR bands. The wide wavelength coverage combined with the parameters of the star 
derived from our spectroscopy provide an excellent base for a robust distance estimate 
based on a Spectral Energy Distribution (SED) fit. 

However, inspection of the available data revealed that the target star is heavily 
saturated in many of the photometric bands covering the optical region 
($\sim0.4\,\mu$m -- 1.0\,$\mu$m), particularly in the automated surveys, such as 
Pan-STARRS (PS1\footnote{https://panstarrs.stsci.edu/}), where the star is saturated 
to such a degree in all survey bands ($g,r,i,z,y$), that no magnitude values are available.

In SDSS (DR13, {\citealt{SDSS-DR13}) measurements of the stellar magnitudes are 
provided, but there are clearly problems in some of the bands. For example, SDSS 
quotes a $z$-band value 2 magnitudes fainter than the quoted $i$-band value, which 
is completely unrealistic for a star with $\teff \sim 5000$ K, as indicated by our spectra. 
Similarly, the $u$-band value is quoted to be 3 magnitudes fainter than the $g$-band, 
which is again not realistic for the type of star we expect and the low extinction in the 
direction of the object (\citealt{Extinction}). 

To rectify this situation, we used our follow-up observations with KeplerCam taken in 
the Sloan $r$, $i$, and $z$ bands to measure the magnitudes ourselves. This was 
achieved in the following fashion: for each band we selected images based on two 
criteria for the stellar profile, (i) FWHM $\le1.8\arcsec$ ($\le2.0\arcsec$ for the 
$r$-band) and (ii) ellipticity $\le 0.1$. These limits were a compromise between the 
actual data quality and the need to have an adequate number of images on the one 
hand; and the requirement that the best images are selected on the other. The above 
cuts typically left us with 20-30\% of the total images in each filter. These images 
were subsequently aligned and co-added, creating three master frames with 
equivalent exposure times of 12, 11, and 23 min for the $r$, $i$, and $z$ band, 
respectively.

We then measured the magnitude of our target in each of the three master images 
through differential photometry using 11 stars in its immediate vicinity which were not 
saturated in the SDSS survey. Our measurements compare well with the spectrally 
matched magnitudes of Tyco 2 stars (\citealt{Pickles2010}) in the $r$ and $i$ bands, 
and are within 0.2 mag for the $z$ band. In the $r$ band the SDSS estimate, the 
spectrally matched value, and our measurement are within 0.1 mag, and the $g$ 
band magnitudes given by SDSS and \cite{Pickles2010} are indistinguishable. For 
this reason, for the SED analysis we used the $u$ and $g$ magnitudes from 
\cite{Pickles2010} and our measurements for the $r$, $i$ and $z$ bands. 

In the $B$ and $V$ bands, independent measurements are available from the 
APASS survey and can also be derived from the Tycho $B_{\rm T}$ and $V_{\rm T}$ 
magnitudes using the \cite{Bessell2000} relations. The TASS photometric catalog 
(\citealt{patch2}) gives an additional independent $V$ band estimate. Most notably, 
the three $V$ band values are consistent to within 0.03 mag, but the $B$ band 
measurements from APASS and Tycho differ by 0.44 mag. We note here, that the 
Tycho measurements have substantial errors ($\sigma_{B_{\rm T}}$ = 0.302 and 
$\sigma_{V_{\rm T}}$ = 0.124) leading to a very uncertain $B-V$ color. Because 
of stated large uncertainties in the Tycho $B_{\rm T}$ and $V_{\rm T}$ magnitudes 
we adopted the $B$ and $V$ APASS values as representative of the true 
brightness of the host star in these bands. We note also that the spectrally matched 
$B$ and $V$ magnitudes from \cite{Pickles2010} are very close to the APASS 
values.

Table\,\ref{tab:StarPars} presents the basic observational --- astrometric, 
photometric and spectroscopic --- parameters of Qatar-6b host star. We used 
these broadband measurements combined with the stellar radius \rstar\ 
derived from the Torres relations to fit an SED using the NextGen library of 
theoretical models and solve for the extinction and distance. We imposed 
a prior of maximum extinction $A_{V} = 0.1$ mag as suggested by the Galactic 
dust reddening maps (\citealt{Extinction}). Figure \ref{fig:Q6bSED} shows that 
the SED of a $\teff \sim 5000$ K main sequence star fits well the broadband 
photometry for a distance $d = 101$ pc and minimal extinction ($A_{V} = 0.05$ 
mag) with a possible exception of the flux in the GALEX NUV band. 

We note that the distance determined from the SED fit compares well with the 
estimated photometric distance. Assuming an absolute magnitude for the host 
star $M_{V}$ = 6.19 (K2V, \citealt{pm2013}), extinction $A_{V} = 0.05$, and 
apparent magnitude $V$ = 11.438$\pm$0.08 (Table\,\ref{tab:StarPars}) yields 
photometric distance $d_{\rm phot}$ = 101-104 pc. 

\begin{figure}
\centering
\includegraphics[width=9.5cm]{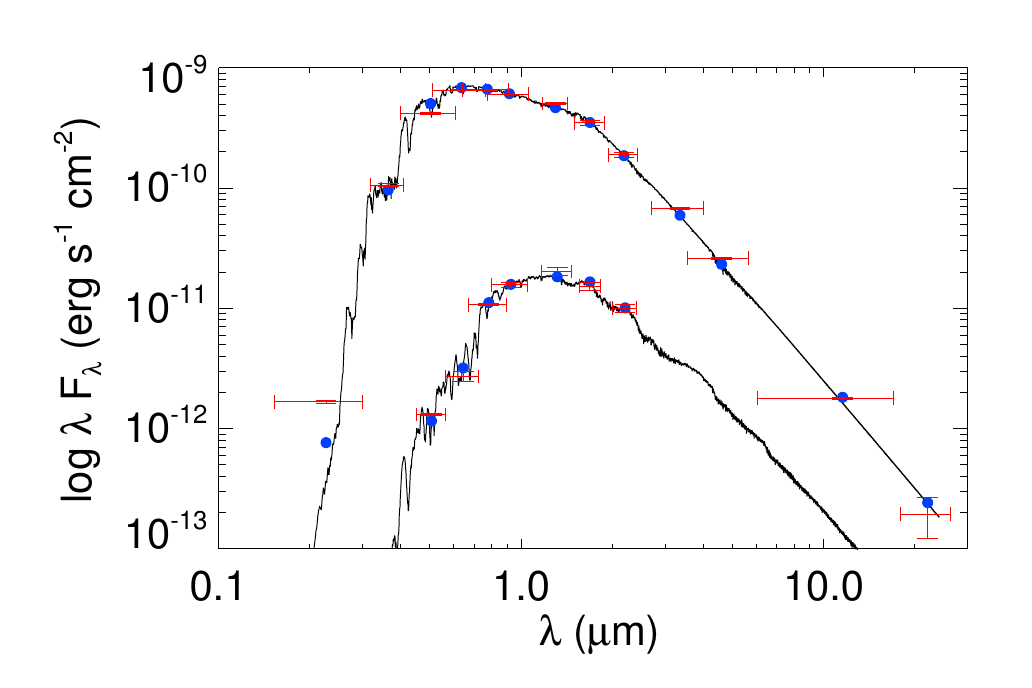}
\caption{{\it Upper curve}: Spectral Energy Distribution (SED) fit for Qatar-6b host 
star. Photometric measurements, summarized in Table \ref{tab:StarPars}, are plotted 
as error bars, where the vertical error bars are the 1\,$\sigma$ uncertainties, whereas 
the horizontal bars denote the effective width of the passbands. The solid curve is 
the best fit SED from the NextGen models where stellar parameters \rstar, \teff, 
\logg, and \feh\, were kept fixed at the values derived from the global fit (Table 
\ref{table:ExoQ6}), and visual extinction ($A_{V}$) and distance ($d$) were 
allowed to vary. {\it Lower curve} is the SED fit for the companion under the 
assumptions described in \ref{subsec:Companion}.}
\label{fig:Q6bSED}
\end{figure}
\noindent

\subsection{The close-by object} \label{subsec:Companion}

Inspection of our KeplerCam follow-up data readily revealed a neighboring object, 
approximately $4\farcs5$ due South from the primary star. This object  is naturally 
blended with the primary in the QES survey images, because of the coarse pixel 
size, but it is clearly visible and detected in all surveys with good spatial resolution 
imaging data (PanSTARRS, SDSS and 2MASS). An SDSS composite image of our 
target star and the neighboring object is shown in Figure\,\ref{fig:Q6bCompanion}.

\begin{figure}
\centering
\includegraphics[width=6.0cm]{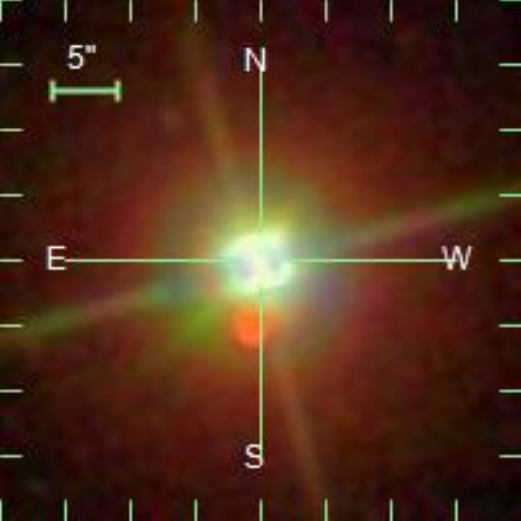}
\caption{SDSS color composite image of Qatar-6b host star and neighboring object.}
\label{fig:Q6bCompanion}
\end{figure}
\noindent

In the SDSS, the object is classified as a galaxy and even a photometric redshift is given 
in the catalog. However, the object is located in the wings of a heavily saturated PSF and 
given the problems with estimating correct magnitudes of the primary star in at least 
some of the survey bands, the quoted values for this neighboring object are also suspect. 

To address this problem we used the master KeplerCam frames, from the follow-up 
observations described above, 
to measure the neighbor/primary flux ratio. We employed two methods: (1) a PSF, 
determined from several isolated stars in close proximity to the target was aligned 
with the primary, scaled and subtracted; (2) a postage-stamp image centered on 
the primary was cut out, flipped along the $y$-axis, then aligned and subtracted 
from the original. Both methods allowed us to isolate the neighboring object and 
measure its flux using aperture photometry. This flux was then compared with the 
flux from the primary measured in matching aperture radii. To avoid substantial 
contamination from residuals from the subtraction we used 3 and 4 pixel aperture 
radius. We  measured flux ratios of $F_{\rm neighbor}/F_{\rm primary}$ = 0.0049, 
0.0146, 0.0266 in the Sloan $r$, $i$ and $z$ bands respectively, corresponding to 
magnitude differences of $\Delta{\rm m}$\,=\,5.80, 4.59, and 3.94 mag. In addition, 
we note that at the limit of KeplerCam images resolution (1\farcs8) we do not detect 
departure from the PSF.

To further investigate the nature of the neighboring object we used the NIRC2 
infrared camera behind the Keck II NGS AO system on 2017 August 3 to obtain 
differential near-infrared photometry of the companion and to examine its shape. 
We operated NIRC2 in its 9.95 mas pixel$^{-1}$ mode (\cite{nirc2scale}) which 
results in a field of view of $\sim$10\arcsec. We used the $J$, $H$ and $K$ filters 
and obtained 2-point dithered images, for sky subtraction, with total exposure times 
of 40\,s in each filter. We calibrated the images with dome flat-fields taken during 
the day. Figure\,\ref{fig:Q6bKeckAO} shows half of the sky subtracted NIRC2 
AO frame (dithered position 1). 

The AO images show, without a doubt, the companion object is a star -- the first 
two Airy rings of the diffraction limited PSF are clearly visible in all three filters. The 
flux ratio was measured separately from the two dithered positions and the results 
were averaged. We note that flux ratios determined from the two dithered positions 
are constant to within the measurements uncertainties for all aperture radii between 
$R_{\rm ap} = 0\farcs3-0\farcs7$. The estimated values in the $J$, $H$ and $K$ 
bands are $F_{\rm neighbor}/F_{\rm primary}$ = 0.0382, 0.0438, 0.0512 respectively, 
corresponding to magnitude differences $\Delta{\rm m}$\,=\,3.54, 3.40, and 3.19 mag.  

\begin{figure}
\centering
\includegraphics[width=6.0cm]{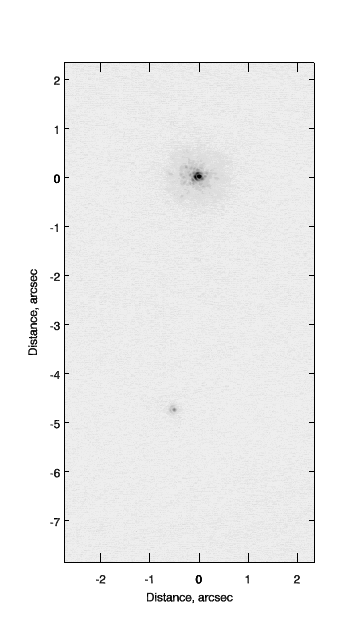}
\caption{NIRC2 Keck AO image of Qatar-6b host star and its neighbor. Shown is 
one half the sky subtracted, flat field corrected dithered position 1. The image is 
stretched by the square root of the intensity to highlight the Airy rings.}
\label{fig:Q6bKeckAO}
\end{figure}
\noindent

From the flux ratio measurements described above we derive the apparent magnitudes 
of the companion object --- 17.02, 15.49, 14.69, 13.26, 13.71, 12.41 in the $r$, $i$, $z$, 
$J$, $H$, and $K$ filters, respectively. The estimated uncertainties are 0.05-0.1 mag 
in the Sloan bands, dominated by the image subtraction residuals, and 0.05-0.08 mag 
in the NIR filters. The observed broad band magnitudes constrain well the shape of the 
SED and indicate the companion is a late M star with $T_{\rm eff} \approx 3200$ K. 
Unfortunately, we do not have a independent handle on the stellar radius that will allow 
us to determine the distance to the companion. If we assume it is a MS star, the 
$\rstar$ can be estimated using the semi-empirical $T_{\rm eff}$-radius relation 
derived by \cite{Mann15}. The nominal value from that relation is $\rstar = 0.25 \rsun$, 
corresponding to a distance to the neighbor $D \approx 90$ pc. There is, however, a
significant uncertainty ($\sim$15\%) associated with the $T_{\rm eff}$-radius relation 
leading to a substantial spread in distances. It is interesting to note that for 
$\rstar = 0.28 \rsun$, i.e., only 1$\sigma$ away from the nominal value, the estimated 
distance to the companion is the same as to the primary, $D \approx 100$ pc 
(see Figure \ref{fig:Q6bSED}) in which case the projected distance between the two 
stars would be $\sim 450$ AU. 

To further test whether the two stars are possibly associated we place them on a 
color-magnitude diagram (CMD) using the {\it Gaia} $G$ (DR1, \citealt{GaiaDR1}) 
and 2MASS $J$ magnitudes. {\it Gaia} separates the two stars cleanly and lists their 
magnitudes as 11.076 and 15.982, respectively. For the $J$-band we use the 2MASS 
magnitude of the primary and our estimate of the companion magnitude based on the 
Keck AO images. Figure \ref{fig:Q6bCMD} shows the CMD for the two stars and a 1 
Gyr theoretical isochrone from the Padova database (\citealt{Padova}). The CMD clearly 
shows our estimates of the primary star parameters and the distance are close to 
the theoretically expected values for a 1 Gyr old star. The companion object, on the 
other hand, projects about 3$\sigma$ away from the theoretical isochrone. At face 
value, the CMD does not support the association scenario; however, given the 
uncertainties and assumptions about the close-by star, we cannot draw any definite 
conclusion. Further investigation of possible association is beyond the scopes of this 
paper, and can be carried out much more reliably once the proper motion of both stars 
is measured by {\it Gaia}.

\begin{figure}
\centering
\includegraphics[width=7.0cm]{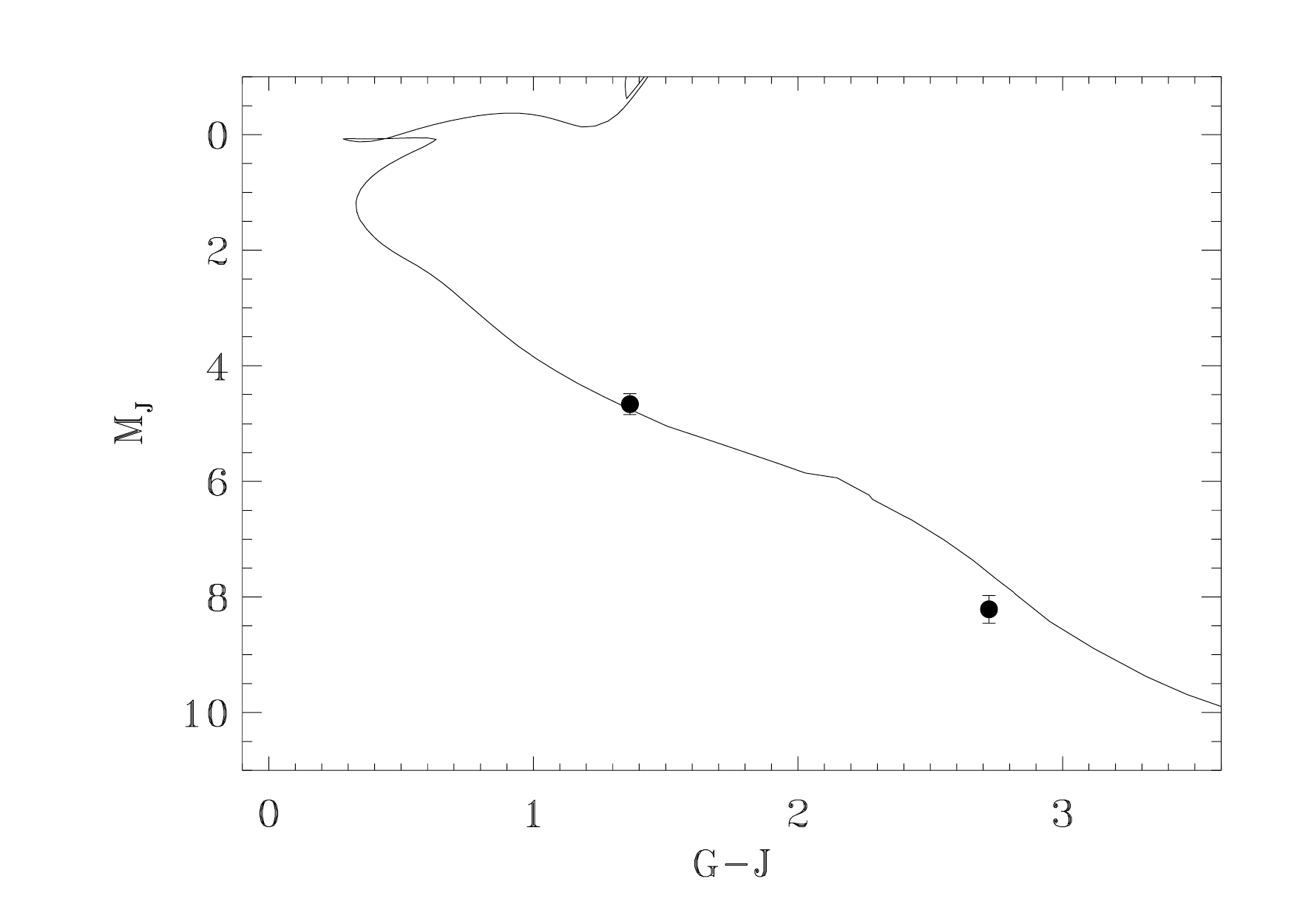}
\caption{Color-magnitude diagram ({\it Gaia} $G-J$, $J$) for Qatar-6b host star and 
its neighbor for a distance $d = 101$ pc. Overplotted is a 1 Gyr isochrone form the 
Padova database (\citealt{Padova}) assuming solar abundances. The error bars are 
dominated by the uncertainty in the distance estimate.}
\label{fig:Q6bCMD}
\end{figure}
\noindent

Eclipsing binaries, both projected and gravitationally bound, are the main source of 
false positive detections in transit surveys. Fortunately, this is not the case for Qatar-6b 
for two reasons. First, being more that 5 mag fainter, the companion object can not 
produce the observed transit depth of $\sim$2\% in the $r$ and $V$ bands. Second, 
the observed RVs, which are measured from the part of the spectrum covering the 
Sloan $g$ and $r$ bands, can only come from the primary star as the companion 
contributes $<1\%$ to the combined light in this wavelength range. Thus, the 
observed RV curve reflects the orbital motion of the primary star only and it is at the 
same period as the transits.
 
\subsection{Orbital period determination} \label{subsec:OrbitalPeriod}

To better determine the transiting planet ephemeris we used the data from the follow-up 
photometric curves. As described in a previous Section, we have observed 6 transits of 
Qatar-6b between January 30 and June 9, 2017, spanning 37 orbital cycles of the system. 
First, all observational time stamps were placed on the BJD$_{TDB}$ system and each 
light curve was rebinned to a uniform cadence of 2 min for the KeplerCam observations 
and 2.5 min for the Meade LX200GPS observations, to reduce the error bars on individual 
points. When binning the errors were propagated assuming data points were uncorrelated 
and we checked that this level of binning did not affect derived parameters. Each individual 
light curve was then fitted with a transit model using {\sc EXOFAST} (\citealt{exofast}). 
We note that {\sc EXOFAST} uses a wavelength dependent quadratic limb darkening 
prescription and uncertainties of the fitting parameters are estimated via Markov Chain 
Monte Carlo (MCMC) minimization. 

\begin{table}
\centering
\caption{Central times of Qatar-6b transits and their uncertainties.}
\label{tab:Tcen}
\begin{tabular}{lrll}
\hline\hline
Transit central time             & Cycle & Filter & Telescope/\\
BJD$_{TDB}$ - 2,400,000  & No. & &Instrument\\
\hline
57784.03257  $\pm$ 0.00068 &  0 & Sloan $i$ & KeplerCam\\
57822.60106  $\pm$ 0.00076 & 11 & $V$ & Meade\\
57836.62569  $\pm$ 0.00048 & 15 & $V$ & Meade\\
57878.70010  $\pm$ 0.00043 & 27 & Sloan $i$ & KeplerCam\\
57892.72446  $\pm$ 0.00036 & 31 & Sloan $r$ & KeplerCam\\
57913.76215  $\pm$ 0.00047 & 37 & Sloan $z$ & KeplerCam\\   
\hline
\end{tabular}
\end{table}

The transit central times $T_{C}$ and their uncertainties were estimated from the model 
fits and we calculated the best ephemeris by fitting a straight line through all the points. 
Figure\,\ref{fig:OrbPeriod} shows the fit and the residuals from the linear ephemeris, and 
the measurements of $T_{C}$ and their uncertainties are summarized in Table\,\ref{tab:Tcen}. 
The final orbital ephemeris is expressed as 

\begin{equation}
T_{C} = 2457784.03270(49) + 3.506195(18)\,E 
\end{equation}

\noindent 
where $E$ is the number of cycles after the reference epoch, which we take to be the 
$y$-intercept of the linear fit, and the numbers in parenthesis denote the uncertainty of 
the last two digits of the preceding coefficient. Here the reference epoch and the period 
are in days on the $BJD_{TDB}$ scale, and their uncertainties correspond to 42\,s, and 
2\,s, respectively. The quality of the fit as measured by $\chi^{2} = 1.11$ indicates that a 
linear ephemeris is a good match to the available measurements given the quality of the 
observations. Thus, an investigation of possible transit timing variations (TTV) would 
require data of much superior quality. 

\begin{figure}
\centering
\includegraphics[width=8.8cm]{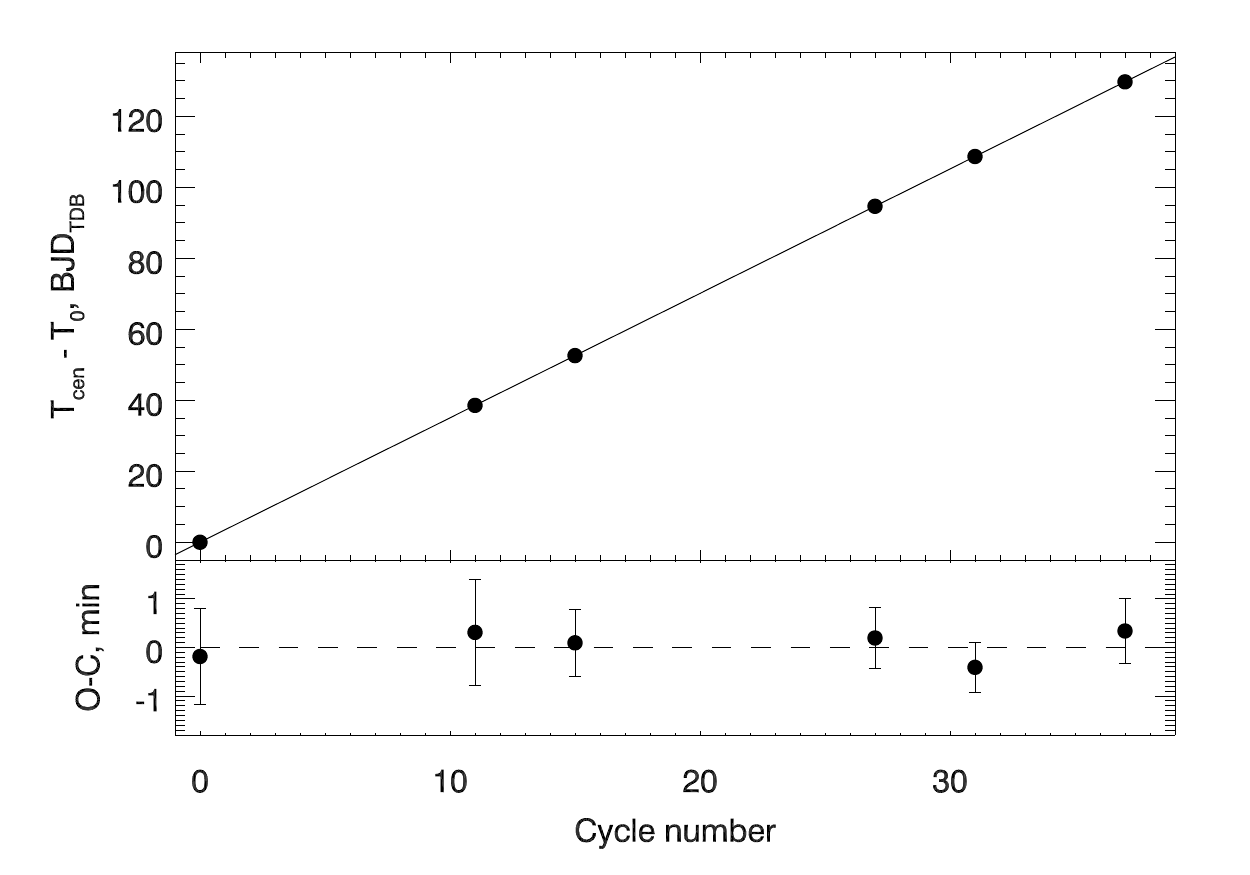}
\caption{The orbital period for Qatar-6b and residuals of central transit times from a 
linear ephemeris.}
\label{fig:OrbPeriod}
\end{figure}

\subsection{Planetary system parameters} \label{subsec:EXOFAST}

To determine the physical parameters of the planetary system we run a global solution of 
the available RV and transit photometric data using version 2 of {\sc EXOFAST} --  a full 
re-write of the original package, designed to simultaneously fit RV and/or transit data for 
multiple planets, transits, and RV sources (\citealt{exofast-v2}). Further description of 
{\sc EXOFASTv2} and similarities and differences with the original version can be found 
in \cite{Eastman2017} and \cite{Rodrigues2017}. One major difference which is important 
in our case is that {\sc EXOFASTv2} uses the YY isochrones (\citealt{YY}) to model the 
star, instead of the Torres relations (\cite{torres}, see also \cite{Eastman2016} for a more 
detailed description). 

The global fit of Qatar-6b includes the RV measurements listed in Table \ref{table:RV}
and the six follow-up photometric light curves shown in Figure \ref{fig:Q6bTR}. The 
stellar parameters (\teff, \logg, \mh) determined from the spectroscopic analysis 
(Section \ref{subsec:SpecPars}) and the orbital period and reference epoch measured 
from the analysis of the transit central times (Section \ref{subsec:OrbitalPeriod}) are 
fed as initial parameters for the global fit. 

The spectroscopically determined stellar parameters and the comparison with the theoretical 
isochrones suggest that the host star is well positioned on the main sequence. Quadratic 
limb darkening coefficients (LDCs) were interpolated from the \cite{LDcoeff} tables for each 
filter and used as initial input. The LDCs were left free to vary, but we imposed a Gaussian 
prior around the interpolated value with an uncertainty of 0.05. An additional Gaussian prior 
was imposed on the stellar age, with a mean value of 1 Gyr and an uncertainty of 0.5 Gyr, 
based on our analysis in Section \ref{subsec:SpecPars}, while an upper-bound uniform prior 
of $A_{V}<0.1$ was imposed on the visual extinction, as suggested by the Galactic dust 
reddening maps (\citealt{Extinction}).

With respect to the eccentricity, using the equations from \cite{leconte} and \cite{jackson}, 
we calculated the tidal circularization time-scale for the system. We used the values from 
Table \ref{table:ExoQ6} of $M_{\star}$, $R_{\star}$, $M_{P}$, $R_{P}$, assuming tidal 
quality factors of $Q_{\star}=10^{6.5}$ and $Q_{P}=10^{5.5}$. The rotation period given 
in Table \ref{table:ExoQ6}, $P_{rot}=12.75$ d, is calculated using the stellar radius from 
our solution, the \vsini\ from the spectra, and assuming the stellar rotation axis and the 
planet orbit are coplanar. The time-scale for circularization is $\tau_{\rm circ}=0.08$ Gyr, 
significantly lower than the estimated age of the host star and, thus, we expect the planet 
orbit to have circularized. Additionally, our RV data is not of high enough quality to allow
investigation of (any potential) small departures from circularity. For these reasons, in 
fitting the Qatar-6b data set, we only considered a circular orbit and kept the eccentricity 
fixed to zero. 

Table \ref{table:ExoQ6} summarizes the physical parameters of the planetary system. 
The best fit for both radial velocity and photometric light curves is coming from the 
{\sc EXOFASTv2} global fit. The Safronov number is not used in the current paper and 
is provided in Table \ref{table:ExoQ6} for completeness, as it may be useful for other 
studies. A plot of the posterior distributions form the MCMC fit for selected stellar and 
planetary parameters is presented in Figure \ref{fig:Q6bCornerPlot} to demonstrate 
the quality of the fit. For the LDCs, the MCMC fits converge very close to the interpolated 
values from \cite{LDcoeff} with typical uncertainty of 0.05.

\begin{table*}
\centering
\caption{Median values and 68\% confidence intervals. We assume 
	$R_{\odot}$=696342.0\,km, $M_{\odot}$=1.98855$\times 10^{30}$\,kg, 
	$R_{\rm J}$ = 69911.0\,km, $M_{\rm J}$=1.8986$\times 10^{27}$\,kg 
	and 1 AU=149597870.7 km.}
\label{table:ExoQ6}
\begin{tabular}{lcc}
\hline
Parameter & Units & Qatar-6b \\
\hline
\multicolumn{2}{l}{Stellar Parameters:} &  \\
    ~~~$M_{*}$\dotfill     &Mass (\msun)\dotfill  & $0.822\pm0.021$ \\
    ~~~$R_{*}$\dotfill      &Radius (\rsun)\dotfill & $0.722\pm0.020$ \\
    ~~~$L_{*}$\dotfill       &Luminosity (\lsun)\dotfill & $0.306\pm0.026$ \\
    ~~~$\rho_*$\dotfill     &Density (g/cm$^{3}$)\dotfill & $3.08\pm0.16$ \\
    ~~~$\log(g_*)$\dotfill &Surface gravity (cgs)\dotfill & $4.636\pm0.014$ \\
    ~~~$\teff$\dotfill        &Effective temperature (K)\dotfill & $5052\pm66$  \\
    ~~~$\feh$\dotfill        &Metallicity\dotfill & $-0.025\pm0.093$ \\
    ~~~$\tau_{\rm YY}$\dotfill   &Age (Gyr)\dotfill & $1.02\pm0.62$   \\
    ~~~$A_{V}$\dotfill    & Extinction (mag)\dotfill & $0.05\pm0.04$ \\
    ~~~$d_{\rm SED}$\dotfill &Distance (pc)\dotfill & $101.5\pm5.6$   \\
\multicolumn{2}{l}{Planetary Parameters:} &  \\
    ~~~$P$\dotfill   &Period (days)\dotfill & $3.506189\pm0.000020$ \\
    ~~~$a$\dotfill   &Semi-major axis (AU)\dotfill & $0.0423\pm0.0004$ \\
    ~~~$M_{P}$\dotfill &Mass (\mj)\dotfill & $0.668\pm0.066$ \\
    ~~~$R_{P}$\dotfill &Radius (\rj)\dotfill & $1.062\pm0.071$ \\
    ~~~$\rho_{P}$\dotfill &Density (g/cm$^{3}$)\dotfill & $0.68\pm0.14$ \\
    ~~~$\log(g_{P})$\dotfill &Surface gravity\dotfill & $3.162\pm0.069$ \\
    ~~~$T_{eq}$\dotfill &Equilibrium Temperature (K)\dotfill & $1006\pm18$ \\
    ~~~$\Theta$\dotfill &Safronov Number\dotfill & $0.064\pm0.007$ \\
    ~~~$e$\dotfill &Eccentricity (fixed)\dotfill & $0$\\
\multicolumn{2}{l}{RV Parameters:} &  \\
    ~~~$K$\dotfill &RV semi-amplitude (m/s)\dotfill & $101.6\pm9.6$\\
    ~~~$\gamma_{\rm rel}$\dotfill &Systemic velocity (m/s)\dotfill & $152.4\pm7.8$ \\
\multicolumn{2}{l}{Primary Transit Parameters:} &  \\
    ~~~$R_{P}/R_{*}$\dotfill &Radius of planet in stellar radii\dotfill & $0.151\pm0.007$ \\
    ~~~$a/R_{*}$\dotfill &Semi-major axis in stellar radii\dotfill & $12.61\pm0.22$\\
    ~~~$i$\dotfill &Inclination (degrees)\dotfill & $86.01\pm0.14$\\
    ~~~$b$\dotfill &Impact Parameter\dotfill & $0.878\pm0.016$ \\
    ~~~$T_{14}$\dotfill &Total duration (days)\dotfill & $0.0662\pm0.0010$\\
\hline
\end{tabular}
\end{table*}
\noindent

\begin{figure*}
\centering
\includegraphics[width=7.0in]{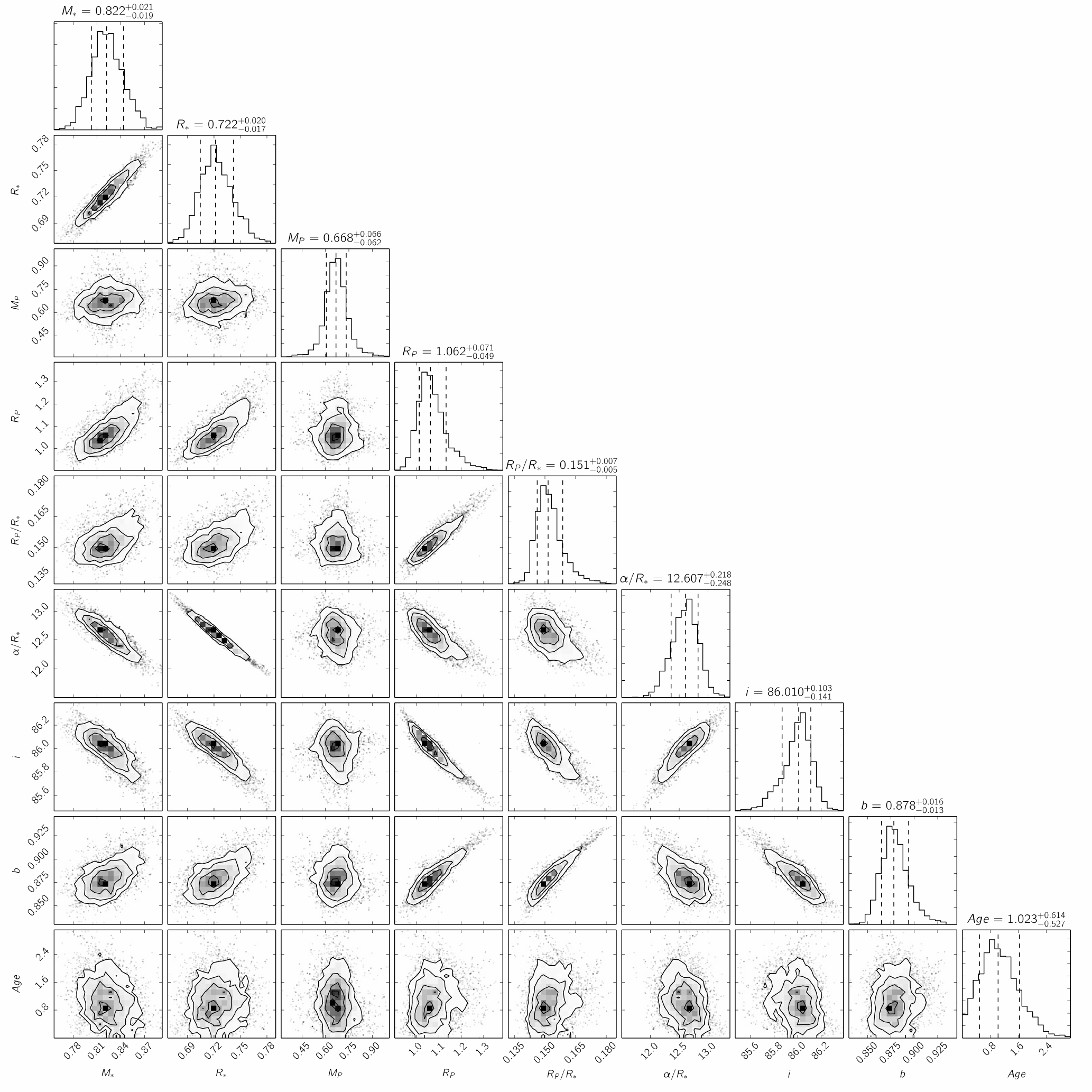}
\caption{Corner plot of the posterior distributions from the MCMC fit for selected 
parameters of the Qatar-6b system.}
\label{fig:Q6bCornerPlot}
\end{figure*}

\subsection{Orbit orientation and limb darkening effects} \label{sec:LD}

The shape of our follow-up transit curves suggested the planet is likely on a grazing transit.
For each transiting planetary system there is a minimum inclination, $i_{gr}$, below which 
the transit is grazing, i.e. part of the planet disk's shadow will always be outside the stellar 
disk even at maximum coverage. This theoretical limit is given by Equation\,\ref{eq:i-gr}

\begin{equation}
\cos i_{gr} = \frac{R_{\star}-R_{\rm p}}{a}
\label{eq:i-gr}
\end{equation}

\noindent
In the case of Qatar-6b, this limit is $i_{gr}=86.16\degr$, using the planetary radius and 
semi-major axis values from Table \ref{table:ExoQ6}. On the other hand, the inclination 
we measure through the global model fit is $i=86.06\degr$. The transit is just grazing 
(considering the errorbars), thus the value of the planet radius we measure from the 
transiting light curves should be very close to its true value (Figure \ref{fig:Q6bOrb}). 
Nevertheless, we need to make clear that, for a grazing transit, the value of the planet 
radius from the model fit should be taken as a lower limit.

\begin{figure}
\centering
\includegraphics[height=8.5cm]{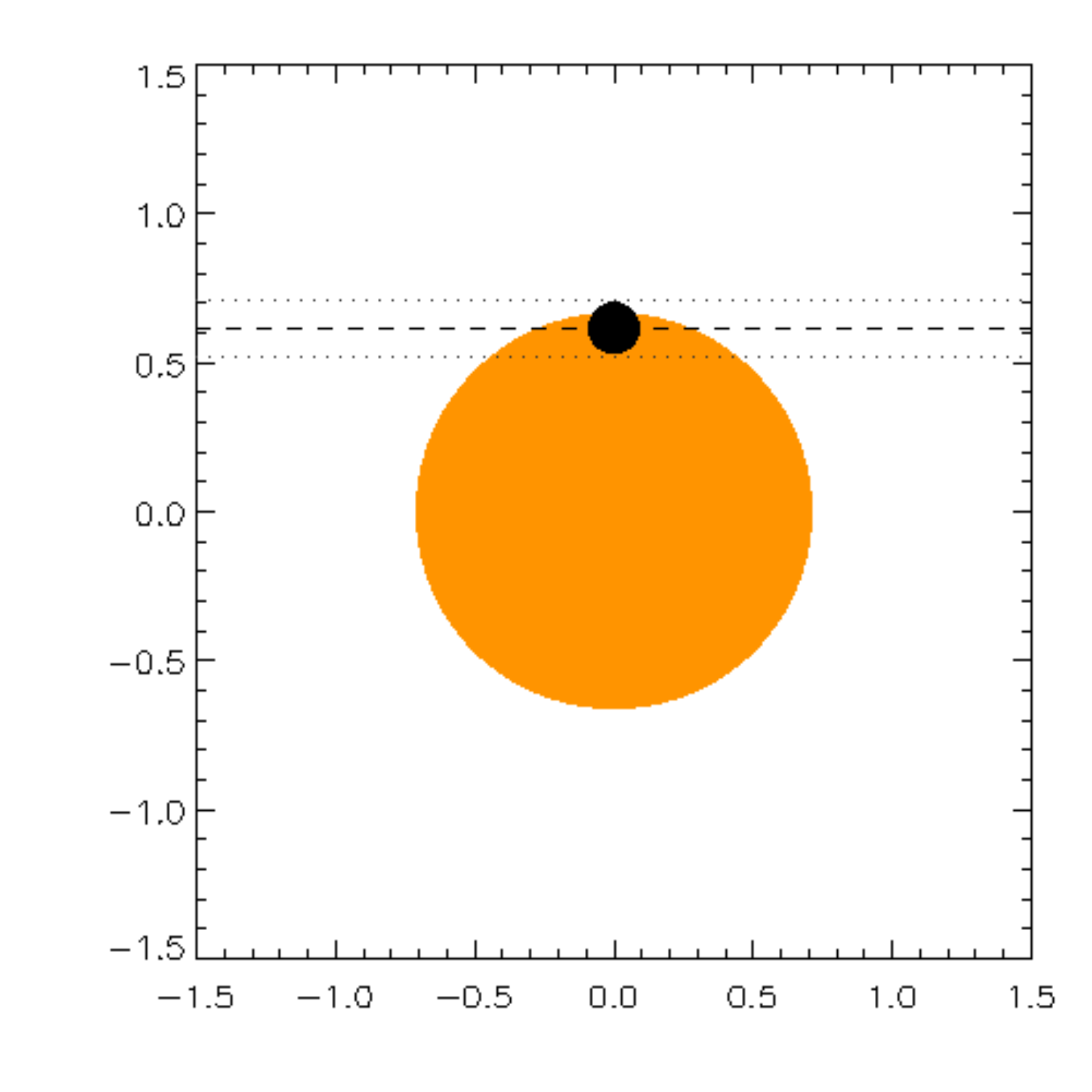}
\caption{Orbital geometry for Qatar-6b, showing the grazing character of the transit. The 
figure is drawn to scale in units of solar radius.}
\label{fig:Q6bOrb}
\end{figure}
\noindent

To a first degree, the shape of a planetary transit light curve is determined by the ratio 
of the planet radius to the stellar radius, the orbit scale and the transit impact parameter. 
Stellar limb darkening (LD), although being a second order effect, also plays an important 
role, as it modifies the shape of the light curve and affects the transit depth. In addition, 
the changes to the shape of the light curve and the depth of the transit are both 
wavelength and impact parameter dependent. In the case of Qatar-6b, our data allow 
us to check if the observed differences in the transit depth conform with the 
expected effects of the stellar limb darkening.

The effects of the LD have been the subject of a number of studies. Here we use the 
formalism of \cite{LD} (see their Appendix B), where the transit depth $\Delta F/F$ 
dependence on the impact parameter and wavelength can be written as

\begin{equation}
\frac{\Delta F}{F} = {k^2}\frac{1-u_{1}(1-\mu) - u_{2}(1-\mu)^2}{1-\frac{u_{1}}{3}-\frac{u_{2}}{6}}
\label{eq:LD}
\end{equation}

\noindent
where $k = R_{\rm p}/R_{\star}$ is the ratio of planet radius to star radius, and $u_{1}$ and 
$u_{2}$ are the linear and quadratic limb darkening coefficients. We also use the common 
notation $\mu = \cos \gamma$, where $\gamma$ is the angle between the normal 
vector of the stellar surface point and the direction to the observer. It is easy to show
that, at the time of maximum light loss, $\mu = \sqrt{1-b^2}$, where $b$ is the impact 
parameter. Keeping in mind that limb darkening coefficients are wavelength dependent, 
Equation\,\ref{eq:LD} describes both the impact parameter and wavelength dependence 
of the transit depth. 

\begin{figure*}
\centering
\includegraphics[width=9.0cm]{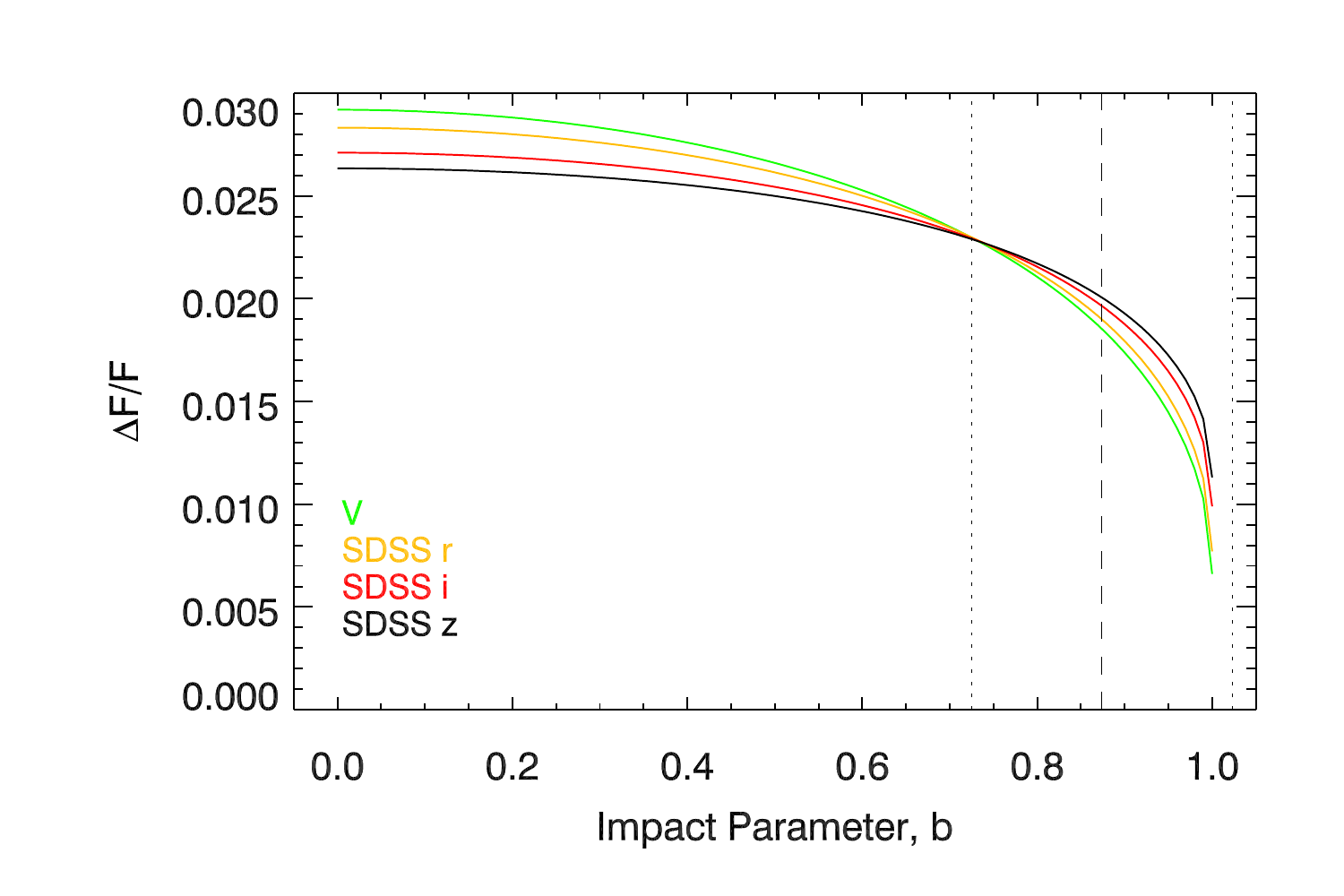}
\includegraphics[width=8.5cm]{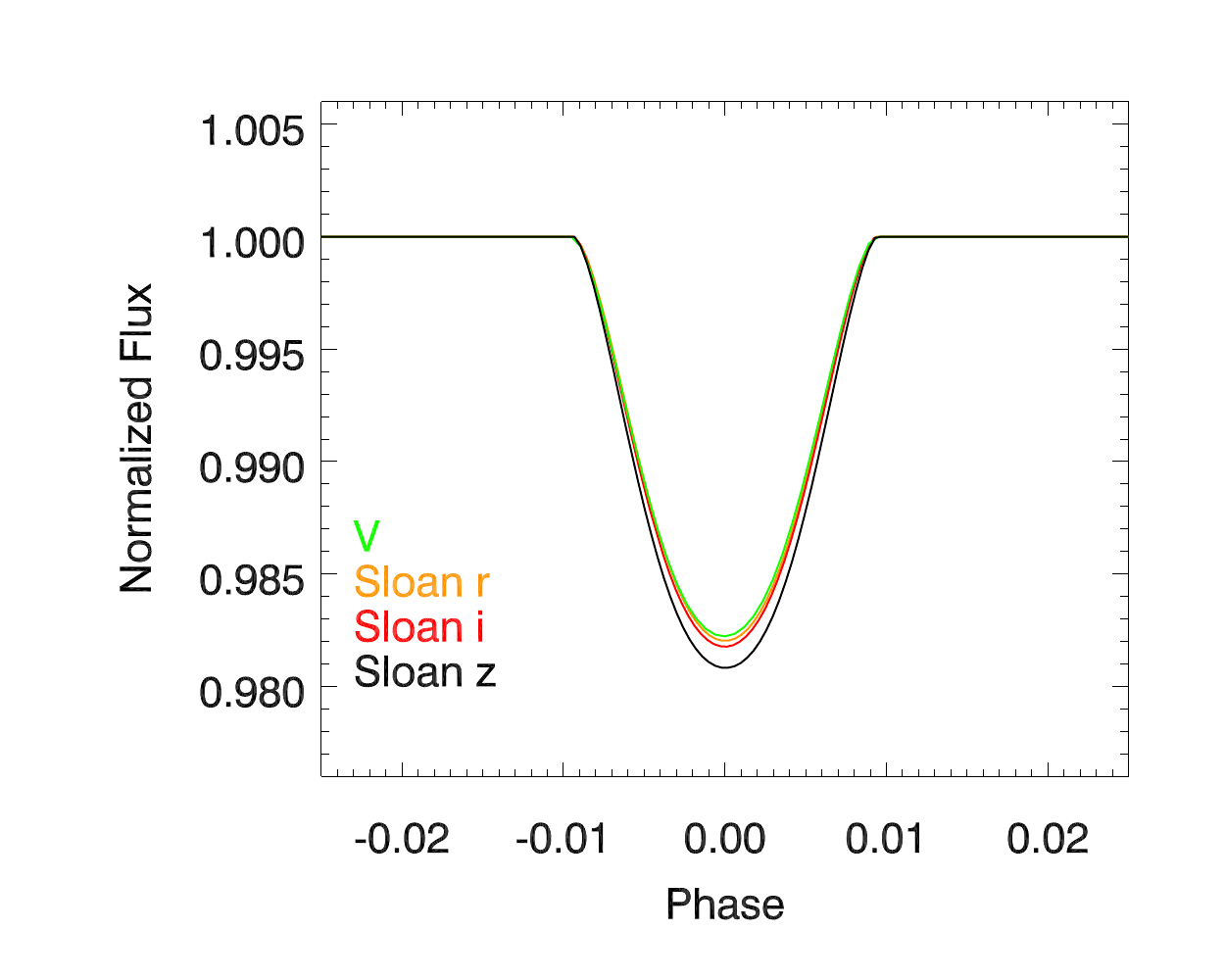}
\caption{{\it Left} panel: Predicted radial profile of the central transit depth $\Delta F/F$ 
as a function of impact parameter (Eq.\ref{eq:LD}). Vertical lines denote the positions of 
the center (dashed) and two edges (dotted) of the planet disk. {\it Right} panel: Observed 
transit depths for Qatar-6b as a function of wavelength. Because of the grazing geometry 
the transit depth is increasing with wavelength as expected.}
\label{fig:Q6bLD}
\end{figure*}

The case of Qatar-6b is demonstrated in Figure \ref{fig:Q6bLD}. The left panel shows the 
expected radial profile of the central transit depth $\Delta F/F$ as a function of the impact 
parameter $b$, as approximated by Eq.\,\ref{eq:LD}. For a given value of $b$, the 
wavelength dependence of $\Delta F/F$ is apparent as different values for curves shown for 
$V$ band (0.55\,$\mu$, green), Sloan $r$ (0.62\,$\mu$, yellow), Sloan $i$ (0.77\,$\mu$, 
red), and Sloan $z$ (0.92\,$\mu$, black). The panel illustrates that for central transits 
(e.g., $b \la 0.5$), the transit depth is expected to decrease with wavelength, while 
for grazing transits ($b \ga 0.8$), it should increase with wavelength. The right panel of 
Figure \ref{fig:Q6bLD} shows what we observe in Qatar-6b and it matches what is 
expected for a grazing transit. For clarity, only the model fits to the filter light curves 
are shown in matching colors as in the left panel. For Qatar-6b we measure central 
transit depth values of 0.0178, 0.0180, 0.0182, and 0.0192 for $V$, Sloan $r$, $i$, 
and $z$ band filters, respectively. These compare reasonably well with the predicted 
values of 0.0185, 0.0190, 0.0196, and 0.0201 from Eq.\ref{eq:LD}. 

\section{Conclusions} \label{sec:conclusions}
 
In this paper, we present the identification of Qatar-6b as a 0.65\mj\, hot Jupiter, orbiting a 
0.817\,\msun, early-K star, with a period of $P_{{\rm orb}}\sim3.5\,{\rm d}$. The age of the 
host star is estimated to be $\tau\sim1\,{\rm Gyr}$, while an SED fit to available and 
measured  multi-band photometry yields a distance of $d = 103$ pc to the system. From 
a global, simultaneous fit to our follow-up photometric and spectroscopic observations we 
measure the planet's radius as $R_{\rm p}$ = 1.04\rj, and demonstrate that the transit is 
grazing, meaning that the $R_{\rm p}$ value should be taken as a lower limit only. We do, 
however, show that the true value of the planetary radius should not be too far off. We also 
investigate the nature of a neighboring object, next to the host star, and conclude that its 
presence is in no way related to the observed periodic variability of Qatar-6b's host and 
does not affect the estimated planetary parameters.

Qatar-6b joins the small family of (nearly) grazing transiting planets. The huge difference in 
numbers between fully and grazing transiting planets is somewhat puzzling. It is true that 
the shape of grazing transits is almost perfectly mimicked by blended eclipsing binaries, so 
the small numbers of grazing exoplanets can (at least partially) be attributed to a selection 
effect, i.e., cases of mistaken false-positive identifications \citep{brown03}. However, 
\citet{oshagh} note that the numbers of grazing exoplanets are still lower than expected and 
propose a physical mechanism where the transit depth of a grazing exoplanet diminishes 
below the detection threshold, under the assumption that the host star harbors a giant polar 
spot. As such, increasing the numbers of known grazing exoplanets is important to better 
understand these detection biases and any potential mechanism behind them.

And finally, planets in grazing transit configurations are especially valuable as they are 
potentially the best targets to look for the presence of additional bodies in their systems
that would lead to detectable variations in the transit impact parameter and duration 
(\citealt{kipping09}, \citealt{kipping10}). These would require high cadence and high 
photometric accuracy observations that could be achieved with large ground based 
telescopes and/or space based facilities.

\section*{Acknowledgements}

This publication is supported by NPRP grant no.\, X-019-1-006 from the Qatar National 
Research Fund (a member of Qatar Foundation). The statements made herein are solely 
the responsibility of the authors. ZT thanks Jason Eastman for his help in installing and  
de-bugging the beta version of {\sc EXOFASTv2} and for the extensive discussions on 
applying different priors and their impact on the global fit. C.B. acknowledges support from 
the Alfred P. Sloan Foundation. Some of the data presented herein were obtained at the 
W. M. Keck Observatory, which is operated as a scientific partnership among the California 
Institute of Technology, the University of California and the National Aeronautics and Space 
Administration. The Observatory was made possible by the generous financial support of 
the W. M. Keck Foundation. AE is supported by the National Natural Science Foundation 
of China (grant no.\,11273051). This work has made use of data from the European Space 
Agency (ESA) mission {\it Gaia} (\href{https://www.cosmos.esa.int/gaia}{https://www.cosmos.esa.int/gaia}), 
processed by the {\it Gaia} Data Processing and Analysis Consortium (DPAC,
\href{https://www.cosmos.esa.int/web/gaia/dpac/consortium}{https://www.cosmos.esa.int/web/gaia/dpac/consortium}). 
Funding for the DPAC has been provided by national institutions, in particular the 
institutions participating in the {\it Gaia} Multilateral Agreement.




\end{document}